\documentclass[aps,prx,reprint,showpacs,superscriptaddress,floatfix]{revtex4-2}
\usepackage[utf8]{inputenc}
\usepackage{graphicx}
\usepackage{booktabs}
\usepackage{amsmath}
\usepackage{tabularx}
\usepackage{hyperref}
\usepackage{orcidlink}
\hypersetup{
colorlinks=true,final=true,
        linkcolor=blue,
        citecolor=blue,
        filecolor=blue,
        urlcolor=blue,
}
\usepackage{bm}
\usepackage[defaultcolor=red]{changes}

\renewcommand{\vec}{\boldsymbol}

\begin{document}
\title{Unraveling Fermi liquid charge and heat transport in RuO$_{2}$}
%Fermi liquid transport in RuO$_{2}$ from dynamical mean-field theory\\
%\AG{Title TBD. In a previous version I proposed: ``RuO$_2$ as a Testbed for Precision Many-Body Electronic Structure: Spectroscopy, 
%Transport and the Role of Hund's Coupling'' perhaps not great either but the current one is a bit bland.}
\author{Harrison LaBollita\orcidlink{0000-0002-6699-8577}}
\email{hlabollita@flatironinstitute.org}
\affiliation{Center for Computational Quantum Physics, Flatiron Institute, New York, New York 10010, USA}             
\author{Fabian B.~Kugler\orcidlink{0000-0002-3108-6607}}
\affiliation{Institute for Theoretical Physics, University of Cologne, 50937 Cologne, Germany}
\author{Antoine Georges\orcidlink{0000-0001-9479-9682}}
\affiliation{Coll{\`e}ge de France, 11 Place Marcelin Berthelot, 75005 Paris, France}
\affiliation{Center for Computational Quantum Physics, Flatiron Institute, New York, New York 10010, USA}             
\affiliation{CPHT, CNRS, {\'E}cole Polytechnique, IP Paris, F-91128 Palaiseau, France}
\affiliation{DQMP, Universit{\'e} de Gen{\`e}ve, 24 Quai Ernest Ansermet, CH-1211 Gen{\`e}ve, Suisse}
\author{Cyrus E. Dreyer\orcidlink{0000-0003-0035-7520}}
\affiliation{Department of Physics and Astronomy,
             Stony Brook University,
             Stony Brook, New York, 11794-3800, USA}
\affiliation{Center for Computational Quantum Physics, Flatiron Institute, New York, New York 10010, USA}

\begin{abstract}
RuO$_2$ is a multi-orbital Fermi liquid for which an array of complementary experimental measurements has recently become available, making it a stringent testbed for quantitative many-body electronic structure theory. Using realistic dynamical mean-field calculations, we compute the spectroscopic and transport properties of RuO$_2$ in the paramagnetic state. We identify RuO$_2$ as a moderately correlated Fermi liquid and obtain (semi-)quantitative agreement with experiment for its optical conductivity, electrical resistivity, and thermopower. This broad agreement across complementary observables establishes a ``precision'' many-body description of RuO$_2$ and allows remaining discrepancies to be identified as meaningful tests of the theory. In particular, the calculated electronic thermal resistivity significantly undershoots the experimentally inferred low-temperature thermal resistivity, raising an important puzzle concerning the relative relaxation of charge and heat currents in the Fermi liquid  regime. We explore low-temperature electron-phonon scattering as a possible mechanism for resolving this discrepancy, while a fully quantitative description remains an open problem. More broadly, our results demonstrate the power of combining precision DFT+DMFT calculations with complementary experimental probes to build and critically test quantitative descriptions of correlated quantum materials.
\end{abstract}

\maketitle

\section{\label{sec:intro}Introduction}
Rutile ruthenium dioxide (RuO$_{2}$) is a metallic transition-metal oxide whose electronic structure has been extensively characterized across multiple experimental probes. Early studies established RuO$_{2}$ as a Pauli-paramagnetic metal with a multiband Fermi surface~\cite{graebner1976}, which has been well-characterized by band theory calculations~\cite{mattheiss1976,yavroksy1996}. More recently, renewed interest in this material has been driven by proposals of altermagnetism. While this question is not yet fully settled, some recent experiments point to a paramagnetic state in the bulk and possible 
altermagnetism in thin films due to epitaxial strain~\cite{berlijn2017,zhu2019, gregory2022,hiraishi2024,Kessler2024,smolyanyuk2024,karube2022,bai2022,liu2024,wu2025,Kiefer2025,Plouff2025Revisiting,Feng2022anomalous,Bose2022Tilted,pawula2024}. Due to this renewed interest, many complementary experimental datasets have recently become available for RuO$_{2}$ using a variety of experimental probes~\cite{HUSSAIN2025417723,li2026explorationaltermagnetismmathrmruo2,Paul2026,lin2025,fedchenko2024,osumi2026,liu2024,graebner1976,wu2025}. Specifically, optics and transport measurements provide clear evidence of Fermi liquid (FL) behavior with coherent quasiparticles~\cite{wenzel2025,ling2026} and characterize its fundamental FL properties. This makes RuO$_{2}$ an appealing testbed for many-body electronic structure theory. While previous work has focused on the general electronic and magnetic properties of RuO$_{2}$~\cite{mattheiss1976,yavroksy1996,park26}, a quantitative many-body description of how electronic correlations reshape the electronic structure and therefore its low-temperature FL transport properties is lacking.

Here, we fill this gap by calculating the many-body electronic structure and transport properties of RuO$_2$ based on the combination of density-functional theory (DFT) and dynamical mean-field theory (DMFT)~\cite{Georges1996,Kotliar2006}. We identify RuO$_2$ as a moderately correlated FL with a high coherence scale. The calculated many-body electronic structure and optical conductivity agree with experiment~\cite{lin2025,fedchenko2024,osumi2026,liu2024}, with dynamical correlation effects playing an essential role in reproducing the low-energy interband transitions~\cite{wenzel2025,park26}. Using this validated many-body electronic structure, we compute within the single-site DMFT approximation the temperature dependence of the electrical conductivity, thermopower, and electronic thermal conductivity. The calculated $T^{2}$ and $T$ coefficients of the electrical resistivity and thermopower in the FL regime agree semi-quantitatively with experiment~\cite{ling2026,pawula2024}, while the corresponding $T^{2}$ coefficient of the thermal resistivity shows a large discrepancy~\cite{ling2026}. The broad agreement across electronic structure and transport observables allows us to identify this heat-versus-charge discrepancy as a meaningful transport puzzle. We explore low-temperature electron-phonon scattering as a possible resolution, while a fully quantitative understanding remains an open problem. More broadly, our work highlights the power of precision many-body calculations combined with complementary experimental probes for developing and critically testing quantitative descriptions of quantum materials.

The rest of this paper is organized as follows. Section~\ref{sec:model} describes our theoretical framework including the crystal structure, definition and methodological approach to the quantum embedding problem, and a recap of the transport definitions used throughout. Section~\ref{sec:results} contains our main results. In Sec.~\ref{sec:structure}, we summarize the basic crystal and electronic structure, as well as the construction of the near Fermi-level Ru-$t_{2g}$ bands for which many-body effects will be considered. In Sec.~\ref{sec:optical}, we present the many-body electronic structure of RuO$_{2}$ and its corresponding optical conductivity. In Sec.~\ref{sec:transport_results}, we present DMFT calculations of the low-temperature transport properties of RuO$_{2}$ compared with experiment. In Sec.~\ref{sec:elph}, we discuss the possible role of electron-phonon contributions to low-temperature charge and heat transport in comparison with recent experiments~\cite{ling2026}. Section~\ref{sec:summary} is a summary and discussion. Appendices contain the DMFT transport data and a description of the electron-phonon model.

\section{\label{sec:model}Theoretical framework}
\subsection{\label{sec:compsettings}Computational settings} 
RuO$_{2}$ crystallizes in the tetragonal $P4_{2}/mnm$ space group giving rise to a primitive cell with two formula units. The occupied Wyckoff positions are 2a (0, 0, 0) for Ru and 4f ($x$, $x$, 0) for O with $x = 0.3056$. We use experimental lattice constants $a=4.419$ \AA{} and $c=3.108$ \AA{}~\cite{Kiefer2025}. Density-functional theory (DFT) calculations were performed with Q\textsc{uantum} ESPRESSO~\cite{qe2009,qe2017} using norm-conserving pseudopotentials parameterized for the Perdew--Burke--Ernzerhof (PBE) exchange correlation functional~\cite{Perdew:1996iq}. The plane-wave basis set size was set with a kinetic energy cutoff of 100 Ry for the wavefunctions. Brillouin zone integration was performed on a $7\times7\times10$ $k$-grid with a Methefessel--Paxton smearing of 0.015 Ry.

We construct a $6\times6$ Wannier Hamiltonian using maximally-localized Wannier functions (MLWFS) as implemented in {\sc Wannier90}~\cite{wannier90} which faithfully interpolates the near Fermi-level Kohn--Sham DFT band structure corresponding to the Ru-$\{x^{2}-y^{2}, xz, yz\}$ states. For this set of localized orbitals, we compute the screened Coulomb interaction within the constrained random phase approximation (cRPA) as implemented in \textsc{CoQui}~\cite{cyeh_crpa} using 300 Kohn-Sham bands, an electronic temperature of $T\sim 1/10$ eV, and $\omega_{\mathrm{max}} = 11$ Ry. We find an intraorbital Coulomb ($U$) of 2.44\,eV and Hund's coupling ($J$) of 0.14\,eV. 

Parameterizing the screened Coulomb interaction in a solid is a consistent challenge when combining DMFT with realistic band structures from DFT. For Ru-based oxides, the Hund's coupling is especially important to quantitatively capture many-body effects in comparison with experiment~\cite{Georges2013,Dang2015}. For example, in SrRuO$_{3}$ and CaRuO$_{3}$, it has been shown that cRPA finds a Hund's coupling that is too small to qualitatively describe experimental phenomenology, such as ferromagnetic ordering~\cite{labollita2026,Dang2015}. In Ref.~\citenum{labollita2026}, we found increasing the Hund's coupling to 0.4 eV was necessary to qualitatively describe the ferromagnetic ordering. Thus, in the following, we compare results using $J=0.2$ eV, i.e., close to the cRPA value, and $J=0.4$ eV, as used in previous studies of ruthenates~\cite{labollita2026}. We fix $U$ to the cRPA value and the interorbital Coulomb term to be $U^{\prime} = U -2J$. 

To describe the quasiparticle and transport properties of RuO$_{2}$, we compute the low-energy interacting electronic structure of RuO$_{2}$ within the single-site DMFT approximation. Starting from the $6\times6$ Wannier Hamiltonian derived from  DFT, we construct a three-orbital quantum impurity model for the Ru-$t_{2g}$ orbitals. Formally, the two Ru sites in the primitive cell correspond to two impurity problems; however, since these sites are equivalent via symmetry, the problem reduces to a single three-orbital impurity model to be solved self-consistently. The one-body part of the impurity Hamiltonian, following from the Wannier Hamiltonian, is augmented with local interactions of the rotationally-invariant Hubbard--Kanamori form, parameterized by the intraorbital Coulomb interaction $U$ and Hund's coupling $J$, with $U^{\prime} = U-2J$. The quantum impurity problem is solved using a continuous-time quantum Monte Carlo solver within the hybridization expansion as implemented in the TRIQS software library~\cite{parcollet_triqs_2015,Seth2016274}. The DMFT self-consistency loop is implemented using TRIQS/ModEST~\cite{modest26} and solved for inverse temperatures $\beta = 10 -200 $ eV$^{-1}$. The Pad\'e analytic continuation method is used to obtain self-energies on the real-frequency axis.

\begin{figure*}
\includegraphics[width=2\columnwidth]{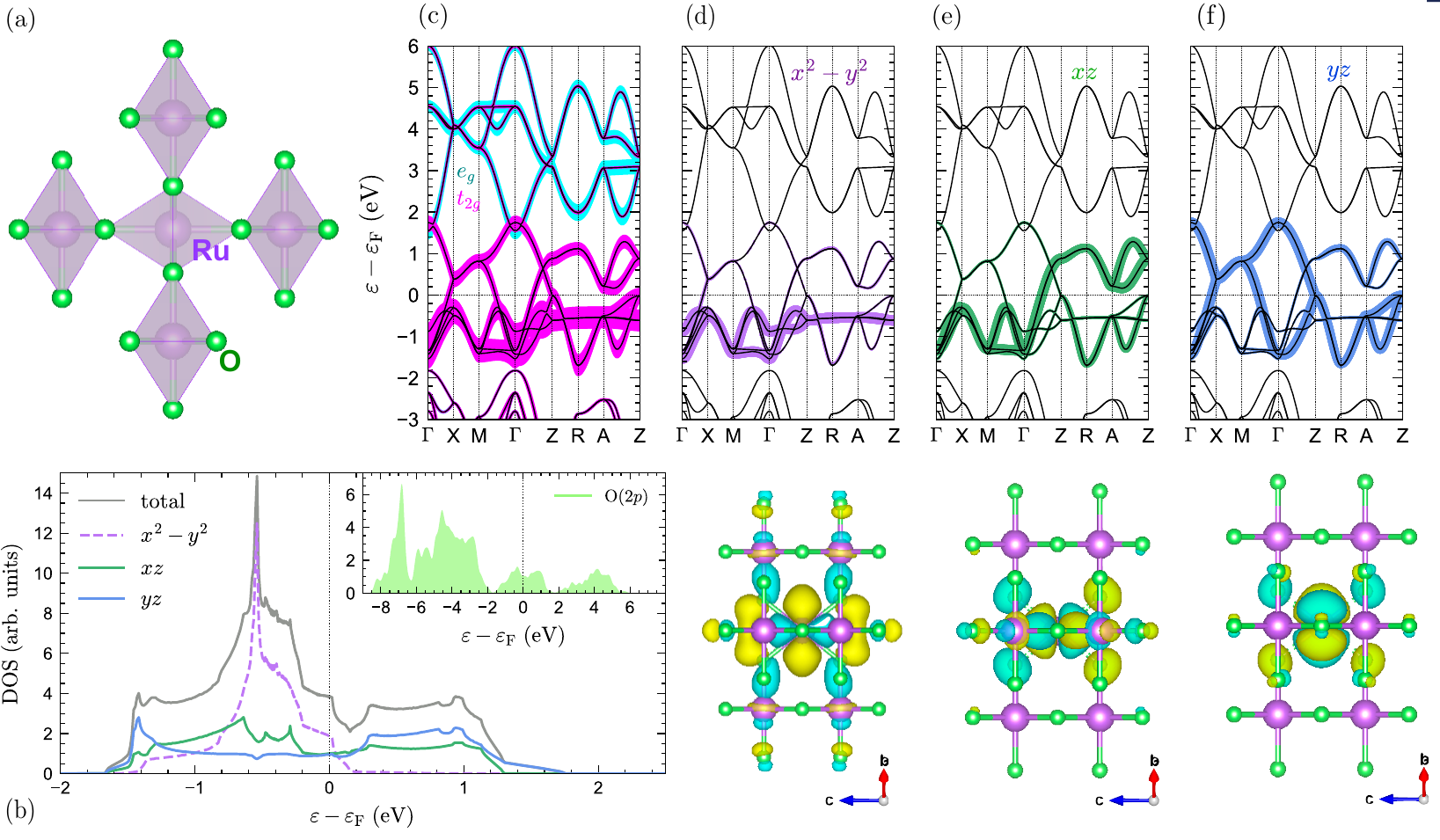}
    \caption{Crystal structure and paramagnetic electronic structure of RuO$_{2}$ within density-functional theory (DFT). (a) Rutile ($P4_{2}/mnm$) crystal structure: Ru and O are denoted as purple and green spheres, respectively. (b) Orbital-resolved Ru-$t_{2g}$ density of states (DOS) around the Fermi level: $x^{2}-y^{2}$ (purpled, dashed), $xz$ (green), and $yz$ (blue). Inset shows O$(2p)$ states in a wide energy window with weight near the Fermi level indicating hybridization with the transition-metal bands. (c) Band structure along high-symmetry lines in the tetragonal Brillouin zone. Colors indicate the dominant orbital character for the Ru-$t_{2g}$ orbitals (orange) and the higher-lying Ru-$e_{g}$ (brown) orbitals. (d,e,f) Orbital weights of the Wannier bands and corresponding constant-value surface of the Wannier orbital for the (d) Ru $x^{2}-y^{2}$, (e) Ru $xz$, and (f) Ru $yz$ orbitals.
    } 
    \label{fig:dft}
\end{figure*}

\subsection{\label{sec:dmft_transport_formalism}Transport formalism}
Within DMFT, neglecting vertex corrections, the optical conductivity simplifies to 
\begin{equation}
\label{eqn:optics}
\sigma_{ab}(\Omega) = 2\pi e^{2}\hbar \int d\omega \, \frac{f(\omega) - f(\omega + \Omega)}{\Omega}\Phi_{ab}(\omega, \Omega),
\end{equation}
where $f(\omega)=(1+e^{\beta\omega})^{-1}$ and $\beta = 1/k_{\mathrm{B}}T$. The transport distribution function $\Phi_{ab}(\omega, \Omega)$ is defined as 
\begin{equation}
\label{eqn:transp}
\Phi_{ab}(\omega, \Omega)= \frac{1}{V} \sum_{\vec{k}}\mathrm{Tr}[v_{\vec{k}}^{a} A_{\vec{k}}(\omega)v_{\vec{k}}^{b}A_{\vec{k}}(\omega+\Omega)],
\end{equation}
where $v_{\vec{k}}^{a}$ are the velocity matrix elements along Cartesian direction $\vec{a}$ and $A_{\vec{k}}(\omega)$ is the interacting spectral function. Note $v_{\vec{k}}^{a}$ and $A_{\vec{k}}(\omega)$ are matrices in the orbital basis. Further details on the transport formalism are provided in the Supplemental Material (SM)~\cite{supp}.

From the dc limit ($\lim_{\Omega\rightarrow0}\sigma_{ab}(\Omega)$), we obtain the direct-current (dc) electrical conductivity. The thermopower $S_{ab}$ and electronic thermal conductivity $[\kappa_{e}]_{ab}$ are obtained from energy-weighted moments of the dc transport distribution $\Phi_{ab}(\omega)\equiv\Phi_{ab}(\omega, 0)$. Explicit definitions for all transport quantities are given in the SM~\cite{supp}.

The dc electrical resistivity is the inverse of the dc electrical conductivity 
\begin{equation}
    \label{eqn:Tel}
\rho^{\text{el}}_{ab} = [\sigma^{-1}]_{ab}.
\end{equation}
To facilitate a direct comparison between the electrical resistivity and the electronic contribution to the thermal resistivity, the latter is often expressed as~\cite{ling2026},
\begin{equation}
\label{eqn:Tth}
[WT]_{ab} = L_{0}T[\kappa_{e}^{-1}]_{ab},
\end{equation}
with $L_{0} = (k_{\rm B}^{2}/e^{2}) \pi^{2}/3$, the Sommerfeld value of the Lorenz ratio. Experimentally, this definition establishes consistent units between the resistivities and ensures the residual $T=0$ thermal resistivity is the same as the electrical resistivity [$WT(0)=\rho^{\mathrm{el}}(0)$].

\section{\label{sec:results}Results}

\subsection{\label{sec:structure}Crystal and electronic structure}
We begin by summarizing the crystal structure and paramagnetic electronic structure of RuO$_{2}$ at the level of DFT. As mentioned above, RuO$_{2}$ has the rutile structure with tetragonal space group $P4_{2}/mnm$~\cite{Kiefer2025} [see Fig.~\ref{fig:dft}(a)]. While the crystal belongs to the tetragonal point group $D_{4h}$, the Ru ions occupy sites with lower $D_{2h}$ symmetry. As a consequence, the local RuO$_{6}$ octahedra deviate from ideal cubic symmetry, and the crystal-field splitting of the Ru($4d$) orbitals is fully nondegenerate. Starting from the ideal octahedra ($O_{h}$) crystal-field picture, where the fivefold $d$ manifold splits into $e_{g}$ and $t_{2g}$ subsets, the reduced $D_{2h}$ symmetry lifts the degeneracy of the $t_{2g}$ manifold, yielding distinct energies for the orbitals of $d_{x^{2}-y^{2}}$, $d_{xz}$, and $d_{yz}$ character.

\begin{figure*}
    \centering
    \includegraphics[width=2\columnwidth]{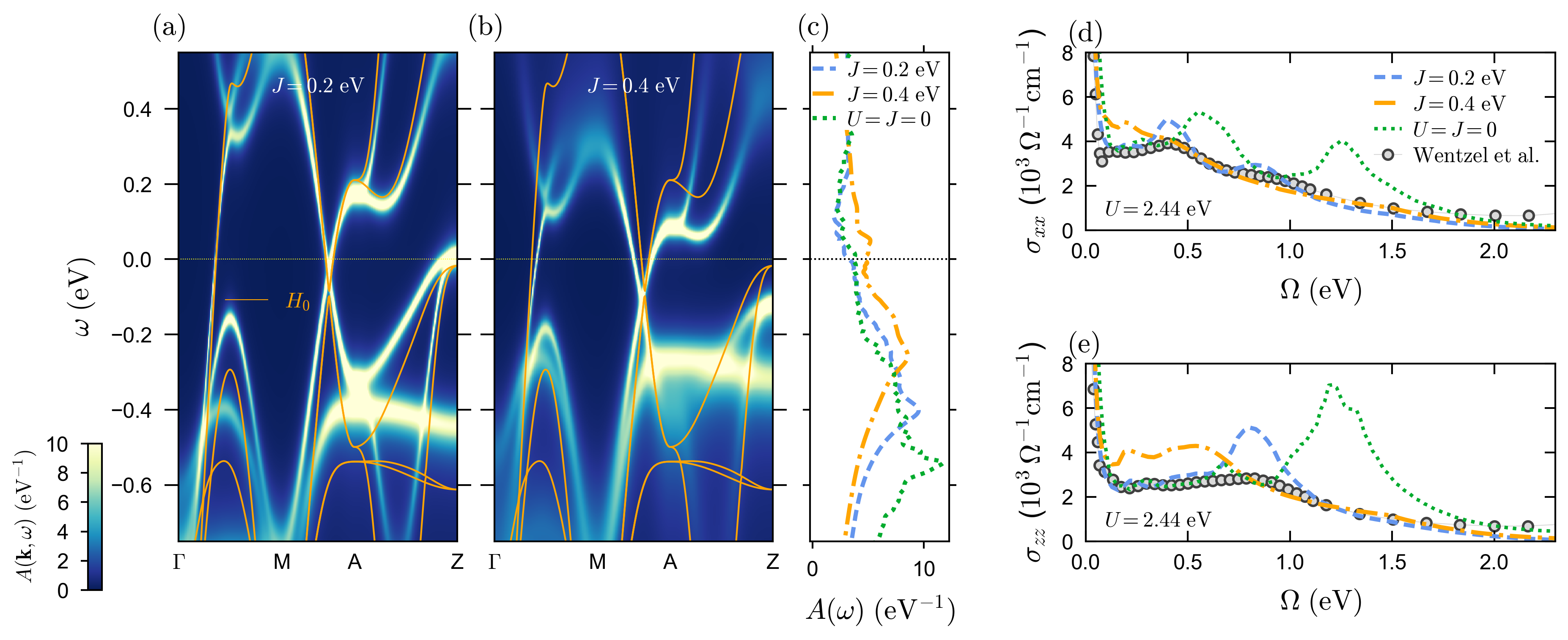}
    \caption{
    Spectroscopic signatures of electronic correlations in RuO$_2$. (a,b) Momentum-resolved spectral function, $A(\mathbf{k},\omega)$, along high-symmetry lines at $T=1/100$ eV ($\simeq116$ K) for $U=2.44$ eV and two values of Hund's coupling: (a) $J=0.2$ eV and (b) $J=0.4$ eV. Thin orange lines show the noninteracting Wannier band structure, $H_0$. (c) Total $k$-integrated spectral function, $A(\omega)$, summed over spin for the same temperature and interaction parameters. (d,e) Optical conductivity for the two values of $J$ along the (d) $xx$ and (e) $zz$ directions. The noninteracting optical conductivity is shown as a dotted gray line for reference. Experimental optical conductivity data are taken from Ref.~\onlinecite{wenzel2025}
    }
    \label{fig:Akw_with_J}
\end{figure*}

Panel (b) of Fig.~\ref{fig:dft} shows the Ru-$t_{2g}$ and O($2p$) density of states (DOS) providing an overview of the electronic structure. The O$(2p)$ states are centered around $-4$ eV and exhibit appreciable hybridization with both Ru-$t_{2g}$ states near the Fermi level and the Ru-$e_{g}$ states at higher energies, as expected from formal valence counting. The low-energy electronic structure is dominated by transition-metal-ligand hybridized states with predominantly Ru-$t_{2g}$ character. The crystal field splitting between the $e_{g}$ and the $t_{2g}$ manifolds is approximately $2.6$ eV, when comparing the band centers. This energy splitting is in good agreement with recent resonant inelastic x-ray scattering (RIXS) measurements~\cite{occhialini2021}.

The DFT band structure reveals six low-lying energy bands of predominantly Ru-$t_{2g}$ character, reflecting the presence of two Ru atoms per primitive unit cell [see Fig.~\ref{fig:dft}(c)]. We interpolate the DFT band structure using maximally-localized Wannier functions obtained by projecting the Kohn--Sham Bloch states within a narrow energy window around the Fermi level onto Wannier orbitals~\cite{wannier90}, as shown in Figs.~\ref{fig:dft}(d)-(f). This procedure yields three orbitals with dominant $x^{2}-y^{2}$, $xz$, or $yz$ character on each Ru site (six orbitals in total). The $x^{2}-y^{2}$ orbital is nearly fully filled and gives rise to relatively flat bands, particularly near the upper $k_{z}$ plane, whereas the $xz$ and $yz$ orbitals form more dispersive bands crossing the Fermi level with comparatively high band velocities. From the resulting Wannier Hamiltonian, we extract local crystal-field energies as $\{\varepsilon_{x^{2}-y^{2}}, \varepsilon_{xz}, \varepsilon_{yz}\} = \{-508,-218,+28 \}$ meV, highlighting the substantial lifting of the $t_{2g}$ degeneracy induced by the local symmetry. With the basic electronic structure of RuO$_{2}$ established, we now turn to the effects of electronic correlations on its electronic structure and transport properties.

\subsection{\label{sec:optical}Electronic correlations revealed in spectroscopy}
 Panels (a) and (b) of Fig.~\ref{fig:Akw_with_J} show as a color plot the momentum-resolved spectral 
function of RuO$_{2}$ at $T = 1/100$ eV  ($\simeq 116$ K) 
for two values of Hund's coupling. The thin (orange) lines denote the band structure from the non-interacting Wannier Hamiltonian ($H_{0}$). At low-energy, the basic band theory dispersions are reproduced with a decrease in the Fermi velocities as the Hund's coupling is increased, which is typical of non-half-filled, multi-orbital metals, where electronic correlations are more sensitive to $J$ than $U$~\cite{Georges2013,Georges2024}. While the dispersive bands with mostly $xz$ and $yz$ character show moderate band renormalization, we find that the states with $x^{2}-y^{2}$ character are most sensitive to Hund's coupling indicated by large energy shifts and substantial broadening of the relatively dispersionless states below the Fermi level along the upper $k_{z}$ plane (A-Z path)~\cite{park26}.  
%\AG{To substantiate the above statement, quote $Z_m$ values - the k-integrated spectra will also help.}
Our results therefore identify these states as a robust spectroscopic probe of local correlation physics in RuO$_{2}$. They call for further ARPES investigations of their energy dispersion and linewidths which will serve as fingerprints of Hund's physics.

Panels (d) and (e) of Fig.~\ref{fig:Akw_with_J} show the optical conductivity of RuO$_{2}$ [Eq.~(\ref{eqn:optics})] for two polarization directions: $xx$ (electric field parallel to the crystallographic $\vec{a}$ axis) and $zz$ (an electric field perpendicular to the $\vec{a}$ axis). 
The optical conductivity obtained at the DFT level is also shown as thin, dashed lines, and is calculated from the bare Wannier Hamiltonian by setting $\Sigma(\omega) = -i\eta$ ($\eta = 10$ meV).
There are two interband transitions, positioned at $\Omega\sim 0.5$~eV and  $\Omega\sim1.25$~eV for the $xx$ direction and at $\Omega \sim 0.6$ eV and $\Omega \sim 1.25$~eV for the $zz$ direction. When compared to recent reflectivity measurements on bulk RuO$_{2}$ crystals, we find that the non-interacting reference electronic structure predicts interband transitions as identified by experiment~\cite{wenzel2025}, but the bare electronic structure dramatically overestimates the intensity and energy position of these transitions. We now compare to the optical conductivity from DMFT at $T = 1/100$ eV ($\simeq 116$ K) for different Hund's coupling parameters. We find local electronic correlations renormalize the location of the interband peaks bringing them closer in line with experiment~\cite{wenzel2025}, specifically the interband transition along the $\vec{c}$ direction corresponding to intra-$t_{2g}$ transitions from the occupied $x^{2}-y^{2}$ states to the unoccupied $xz/yz$ states. 

\begin{table*}
\centering
\begin{tabular*}{2\columnwidth}{c@{\extracolsep{\fill}}cccccc} 
\hline
$J$ & $\gamma$   & $S_{zz}/T$  & $A_{zz}$ & $A_{zz}/\gamma^{2}$ & $B_{zz}$ & $B_{zz}/A_{zz}$\\
(eV)    & ($\mathrm{mJ}\,{\rm mol}^{-1}\,{\rm K}^{-2}$) & $(\mu{\rm VK}^{-2})$ & ($10^{-5}\,\mu\Omega\,{\rm cm}\,{\rm K}^{-2}$) & ($\mu\Omega\,\mathrm{cm}\,\mathrm{mol}^{2}\,\mathrm{K}^{2}\mathrm{J}^{-2}$) & ($10^{-5}\,\mu\Omega\,{\rm cm}\,{\rm K}^{-2}$) & \\
\hline
0.2 & 5.57  & 0.040 & 2.44 & 0.79 & 3.61  & 1.48 \\
%0.3 & 7.14  & X & 3.25 & 0.64 & 5.26  & 1.62 \\
0.4 & 9.73  & 0.150 & 7.79 & 0.82 & 12.25 & 1.57 \\
\hline
Exp. & 5.32$\pm$0.3~\cite{paul2025} & 0.165~\cite{pawula2024}  & 7 $\pm$ 2~\cite{paul2025,ling2026}  & 2.10 & 26 $\pm$ 3~\cite{ling2026}    &  3.70 \\ 
\hline
\end{tabular*}
\caption{Calculated thermodynamic and transport coefficients of RuO$_{2}$ from DMFT for different interaction Hamiltonians parameterized by the Hund's coupling $J$ (with $U = 2.44$ eV). $\gamma$ is the Sommerfeld coefficient corresponding to the linear-in-temperature specific heat coefficient ($C = \gamma T + \cdots$). $S_{zz}$ is the Seebeck coefficient, $A_{zz}$ is the quadratic-in-temperature electrical resistivity coefficient, $B_{zz}$ is the quadratic-in-temperature thermal resistivity coefficient, $A_{zz}/\gamma^{2}$ is the Kadowaki--Woods ratio. See the main text for more details. Experimental data is obtained from Refs.~\cite{ling2026,pawula2024,paul2025}. }
\label{tab:numerics}
\end{table*}

\begin{figure*}
    \centering
    \includegraphics[width=2\columnwidth]{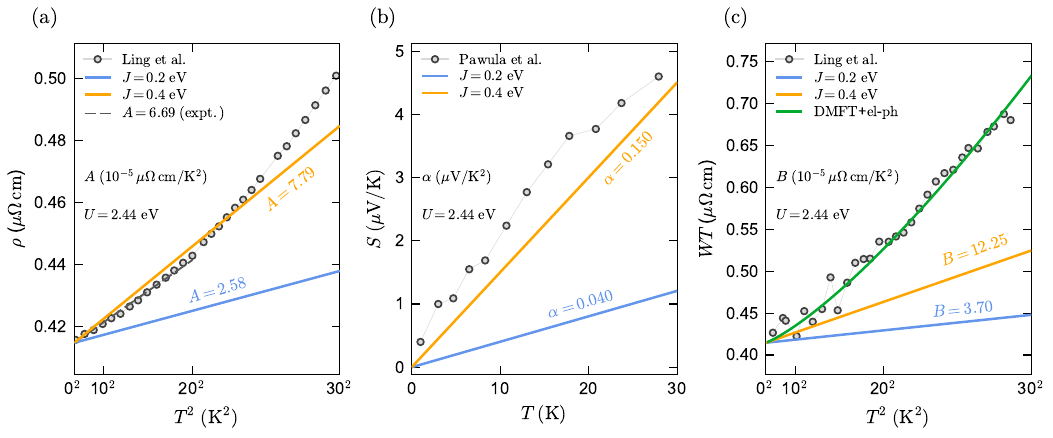}
    \caption{Low-temperature transport properties of RuO$2$ from DMFT compared with experiment~\cite{ling2026,pawula2024} along the $zz$ direction for two values of Hund's coupling $J$, with $U=2.44$ eV fixed. The Fermi liquid coefficients $A$, $\alpha$, and $B$ are obtained by fitting the calculated temperature dependence to $\rho(T)=AT^2$, $S(T)=\alpha T$, and $WT(T)=BT^2$, respectively. (a) Electrical resistivity compared with the experimental data of Ref.~\onlinecite{ling2026}. (b) Thermopower (Seebeck coefficient) compared with digitized experimental data from Ref.~\onlinecite{pawula2024}. (c) Electronic thermal resistivity compared with the experimental data of Ref.~\onlinecite{ling2026}. The DMFT+el-ph curve combines the $J=0.4$ eV DMFT result with the phenomenological Bloch model for electron-phonon contributions to electronic heat transport described in the text.}
    \label{fig:exp_transport}
\end{figure*}

While the comparison of the experimental optical spectra to the DMFT calculations (with different interaction parameters) does not allow to unambiguously discriminate between different values of the Hund coupling, 
it clearly indicates that dynamical correlations are essential for a quantitative comparison. Thus, we can consider RuO$_{2}$ as a moderately correlated, multi-orbital metal. Importantly, by including orbital-dependent local self-energies, $\Sigma_m(\omega)$ obtained with single-site DMFT, we are able to qualitatively and (semi-)quantitatively reproduce the optical data from Ref.~\onlinecite{wenzel2025}.
With our correlated electronic structure validated against spectroscopic experiments, we now turn to the dc limit and focus on the temperature-dependent transport properties of RuO$_{2}$.

\subsection{\label{sec:transport_results}Electronic correlations revealed in transport}
\subsubsection{\label{sec:elres}Electrical resistivity}
The dc resistivity is readily obtained from the optical conductivity as $\rho_{\alpha\alpha}=[\sigma^{-1}(\Omega=0)]_{\alpha\alpha}$.
In a FL, its temperature dependence is expected to be quadratic, $\rho = AT^{2}$, where $A$ is an intrinsic, material-specific quantity~\cite{kugler2026,labollita2026}. We calculate $A_{\rm DMFT}$ by computing the temperature dependence of the electrical resistivity and fitting the DMFT data to $\rho(T) = A_{\rm DMFT}T^{2}$. The raw DMFT calculations are shown in Appendix~\ref{app:extract}. Our calculated quadratic-in-temperature $A$ coefficients are shown in panel (a) of Fig.~\ref{fig:exp_transport} for two values of Hund's coupling $J$. Note that we only quote the values of the resistivity along the $zz$ direction. This is because, despite RuO$_{2}$ being tetragonal ($a= b \neq c$), which allows for anisotropic transport, the dc electrical resistivity is (approximately) isotropic, in agreement with experiment~\cite{Ryden1970} (see Appendix~\ref{app:extract}). The resulting $A$ coefficients for out-of-plane transport are summarized in Table~\ref{tab:numerics}. We find that increasing the Hund's coupling increases the resistivity prefactor $A$. RuO$_2$ remains a coherent FL throughout the parameter range studied, but the stronger low-energy correlations generated by increasing the Hund's coupling $J$ lead to a larger quadratic scattering rate and hence to a larger $T^{2}$ resistivity coefficient.

Recently, the $A$ coefficient of RuO$_{2}$ was characterized in single crystals along the crystallographic $\vec{c}$ direction~\cite{ling2026}. Experimentally, the dc electrical resistivity of RuO$_{2}$ indicates a clear FL regime below $T\sim 20$ K, with $\rho = \rho_{0} + AT^{2}$ attributed to FL electron-electron scattering~\cite{ling2026}. From the transport data in Ref.~\onlinecite{ling2026} the quadratic coefficient $A$ was determined to be $6.6-7.1\times10^{-5}\;\mu\Omega\,\mathrm{cm}/\mathrm{K}^{2}$, ~\cite{ling2026} which falls within the range $5-10\times10^{-5}\;\mu\Omega\,\mathrm{cm}/\mathrm{K}^{2}$ established throughout previous literature~\cite{Ryden1970,Huang1982,wenzel2025,paul2025,Paul2026}. 
Our calculations within the DMFT approximation yield an $A$-coefficient of the correct order of magnitude as compared to experiments, and for larger values of the Hund's coupling reproduce the experimental value quantitatively.

However, agreement in the absolute value of $A$ alone is not a complete test of our calculation. The so-called Kadowaki--Woods ratio (KWR) defined as $A/\gamma^{2}$ is often used to characterize correlated metals as it relates the scattering of quasiparticles ($A$) to their density of states ($\gamma$)~\cite{rice1968, Kadowaki1986, Hussey2005}. Here, $\gamma$ is the linear-in-temperature coefficient of the specific heat and can be obtained from $\gamma = 2 \pi^{2}k_{\mathrm{B}}^{2}/3\sum_{m}A_{m}(0)/Z_{m}$ where $A_{m}(0)$ is the spectral weight at the Fermi energy and $Z_{m}$ is the quasiparticle weight of orbital $m$, respectively. We tabulate our calculated KWRs using our computed $A$ coefficient and specific-heat coefficient in Table~\ref{tab:numerics}. Overall, we find $A/\gamma^{2} = 0.7-0.8\; \mu\Omega\,\mathrm{cm}\,\mathrm{mol}^{2}\,\mathrm{K}^{2}/\mathrm{J}^{2}$, which underestimates the experimental value of $2.1\; \mu\Omega\,\mathrm{cm}\,\mathrm{mol}^{2}\,\mathrm{K}^{2}/\mathrm{J}^{2}$.

Hence we find that, interestingly, the experimentally reported $A$ coefficient can be reproduced quantitatively 
for larger Hund's coupling while a smaller value of that coupling is required to reproduce the experimental  Sommerfeld coefficient (see Table~\ref{tab:numerics}). Correspondingly, we find that the KWR is rather insensitive to the Hund's coupling  and that the calculated value is underestimated as compared to experiments. Altogether, the KWR ratio we obtain is consistent with RuO$_2$ being a moderately correlated metal that is significantly smaller than that of strongly correlated transition-metal oxides. %\hl{to check} \CD{Good point, the $J=0.4$ results put it right on the "Rice" line which contains elemental metals, etc.}

\subsubsection{\label{sec:seebeck}Seebeck coefficient}
The thermopower (Seebeck coefficient) provides complementary information to the dc electrical resistivity because it is particularly sensitive to particle-hole asymmetry near the Fermi level. In the Fermi liquid regime, the thermopower is linear in temperature, $S=\alpha T$. Figure~\ref{fig:exp_transport}(b) shows the calculated thermopower for the two values of Hund's coupling considered here. Increasing $J$ enhances the magnitude of the Seebeck coefficient, reflecting an increased particle-hole asymmetry of the low-energy electronic structure and larger effecitve mass.
We discuss this particle-hole asymmetry by analyzing the transport function in more detail in the Appendix~\ref{app:extract}. We compare our calculations with recent measurements of bulk RuO$_2$ from Ref.~\cite{pawula2024}, with the corresponding linear coefficients $\alpha$ summarized in Table~\ref{tab:numerics}. The larger value of $J$ yields the closest agreement with the measured low-temperature thermopower. Together with the agreement found for the electronic structure, optical conductivity, and electrical resistivity, the thermopower provides an additional independent validation of our many-body description of RuO$_2$. 

\subsubsection{\label{sec:th_rel}Thermal resistivity}
We now turn to heat transport. Panel (c) of Fig.~\ref{fig:exp_transport} shows the calculated thermal resistivity of RuO$_{2}$ from DMFT using Eq.~\eqref{eqn:Tth}. 
We find a quadratic-in-temperature dependence of thermal resistivity and compute the $B$ coefficient in a similar manner as the $A$ coefficient.  Overall, we find that the magnitude of the quadratic coefficient is larger for thermal than charge transport. For a single-band FL and assuming a local self-energy, one can show that $B/A = \frac{\pi^{2}}{3}\frac{1}{12-\pi^2} \approx 1.5442$~(see Ref.~\onlinecite{herring67}). Taking the ratio of our computed quadratic coefficients, we find $B/A \approx 1.4-1.6$, in the same range as this analytical estimate for a single-band system. Deviations from this value can be expected from self-energy contributions beyond the lowest-order quadratic FL form of the self-energy. 

The electronic contribution to the thermal resistivity is experimentally accessible and has recently been measured in bulk single crystals of RuO$_{2}$~\cite{ling2026}. The comparison with experiment is shown in panel (c) of Fig.~\ref{fig:exp_transport}. Interestingly, the experimentally determined $B$ coefficient is larger than the DMFT value by a factor of $2$ to $7$ depending on the choice of interaction parameters (see Table~\ref{tab:numerics}). Experimentally the reported $B/A$ ratio is in the range  $\sim3$--$4$ for RuO$_{2}$~\cite{ling2026}, which is almost a factor of two larger than both our DMFT calculation and the analytical value for a single-band FL with a local self-energy. 

Thus, the theory-experiment comparison in Fig.~\ref{fig:exp_transport}(c) raises a ``transport puzzle'' in RuO$_{2}$ and possibly other metals as mentioned in Ref.~\onlinecite{ling2026}. This discrepancy may reflect limitations of single-site DMFT when describing heat transport. However, before 
invoking effects beyond single-site DMFT such as vertex corrections, non-local self-energies, etc., we consider another physical contribution absent from our calculations: electron–phonon scattering, which as we now show provides a natural explanation for the disagreement.

\subsection{\label{sec:elph}Electron-electron versus electron-phonon contributions to electronic thermal resistivity}
In this section, we present a possible explanation for the discrepancy between the theoretical $B$ coefficient and the one extracted from experiment [see Fig.~\ref{fig:exp_transport}(c)] by considering the contribution of electron-phonon (el-ph) scattering to the electronic thermal resistivity. At low temperatures, this contribution scales like $T^3$ making it challenging to disentangle from the $T^2$ el-el contribution~\cite{Ziman1960}. This is in contrast to the case of electrical resistivity, where the low-$T$ el-el and el-ph scattering contributions scale like $T^2$ and $T^5$, respectively.

To explore this possibility, we employ the Bloch model  for the el-ph thermal resistivity given in Eq.~(\ref{eq:bloch_thermal}). This model assumes a Debye phonon characterized by Debye temperature $\Theta$ and a spherical Fermi surface with dimensionless radius $r$ normalized by the size of the Brillouin zone. The model contains one additional fitting parameter $\rho_\Theta$ (with dimensions of resistivity) and is material-specific. $\Theta$ and $\rho_\Theta$ can be determined from fitting the experimental $\rho(T)$ to
\begin{equation}
\label{eq:rho_fit}
    \rho(T)=\rho_0+AT^2+\rho_{\text{el-ph}}(T)
\end{equation}
where $\rho_{\text{el-ph}}(T)$ is given by the equivalent Bloch model for electrical resistivity [Eq.~(\ref{eq:bloch_electrical})]. We show in Appendix~\ref{app:bloch} Fig.~\ref{fig:elph}(a) that this provides a good fit to the data in Ref.~\cite{ling2026} over the whole temperature range using our DFT+DMFT-calculated ($J=0.4$ eV) $A_{\rm DMFT}=7.79\times10^{-5}\,\mu\Omega\,{\rm cm}\,{\rm K}^{-2}$.

We then fit the thermal resistivity from Ref.~\cite{ling2026} to
\begin{equation}
\label{eq:WT_fit}
    WT(T)=\rho_0+BT^2+[WT]_{\text{el-ph}}(T),
\end{equation}
where $\rho_0$ is taken from experiment and $B_{\rm DMFT}=12.25\times10^{-5}\,\mu\Omega\,{\rm cm}\,{\rm K}^{-2}$ is taken from our $J=0.4$ eV DFT+DMFT calculations. Thus the only adjustable parameter is $r$ in $[WT]_{\text{el-ph}}$ [see Eq.~(\ref{eq:bloch_thermal})].

Despite the simplicity of the model for the el-ph contribution, we can see from Fig.~\ref{fig:exp_transport}(c) that Eq.~(\ref{eq:WT_fit}) gives an excellent fit of the experimental data. This suggests that our calculated $B$ coefficient may not be inconsistent with the experimental data in Ref.~\cite{ling2026}, and underscores the challenge of separating el-el and el-ph contributions to the electronic thermal transport. 

\section{\label{sec:summary}Summary and Discussion}

Using calculations based on the DFT+DMFT formalism, we have studied the spectroscopic and low-temperature transport properties of RuO$_{2}$. Our calculations identify RuO$_{2}$ as a moderately correlated, multi-orbital FL with a high coherence scale. The resulting spectroscopic and Fermi liquid transport coefficients are in semi-quantitative agreement with an array of experimental data. Our results clearly show that dynamical correlations capture the dominant interaction effects governing both the electronic structure and several of its low-temperature transport properties, making it an attractive and useful materials platform for quantitative tests of many-body electronic structure theory.

We note that a definitive characterization of the screened Coulomb interaction proved difficult despite having an array of high-quality experimental datasets. Specifically, no single choice of Hund's coupling reproduced the optical spectroscopy, low-temperature transport and Sommerfeld coefficient simultaneously, although several of these experiments point to a value of order $\sim 0.4$~eV similar to other ruthenates. This tension highlights a key challenge for ``DFT+'' methods and clearly calls for future work. The recent self-consistent GW+EDMFT framework~\cite{yeh2026} represents an exciting avenue to determine screened Coulomb interactions in solids, in the case of ruthenates, this framework would provide an \textit{ab-initio} estimate of $J$ from first-principles beyond cRPA.

Beyond the potential uncertainty in interaction parameters, our calculations expose several robust discrepancies that provide more stringent tests of the theory. In particular, the KWR remains smaller than the experimentally reported value and is relatively insensitive to variations in the interaction parameters. More strikingly, the calculated quadratic coefficient of the electronic thermal resistivity substantially underestimates the experimentally inferred value~\cite{ling2026}. The latter discrepancy raises an intriguing puzzle concerning the relationship between charge and heat transport in the FL regime of RuO$_{2}$. Interestingly, we show that including the effects of electron-phonon scattering in the experimental analysis can largely explain the theory-experiment discrepancy. A quantitative resolution to this transport puzzle requires improvement on both sides of the comparison: a more complete treatment of electron-electron transport and a microscopic understanding of electron-phonon contributions to the electronic heat current.

More broadly, RuO$_2$ provides a well-constrained setting in which spectroscopic, thermodynamic, electrical, and thermal probes can be compared within a common many-body framework. By performing ``precision'' DFT+DMFT calculations and benchmarking against multiple complementary experimental observables, we establish a quantitatively consistent description of much of the electronic structure and transport phenomenology of RuO$_2$. This level of quantitative agreement is what allows the remaining discrepancies, particularly in thermal transport, to emerge as well-defined puzzles rather than obvious limitations of the electronic structure calculation. More generally, our work highlights the power of combining precision many-body calculations with complementary experimental probes to build and critically test a unified description of quantum materials.

\begin{acknowledgements}
We thank K. Behnia for fruitful discussions and Y. Ling for sharing their experimental resistivity data from Ref.~\cite{ling2026}. H.L. thanks Ina Park for useful discussions and a critical reading of the manuscript. F.B.K.\ acknowledges funding from the Ministerium f\"ur Kultur und Wissenschaft des Landes Nordrhein-Westfalen (NRW-R\"uckkehrprogramm). C.E.D.\ acknowledges support from the National Science Foundation under Grant No.~DMR-2237674.
The Flatiron Institute is a division of the Simons Foundation. 
\end{acknowledgements}

\section*{Data availability}
The data generated in this study will be made publicly available upon publication.

\appendix

\begin{figure*}
    \centering
    \includegraphics[width=2\columnwidth]{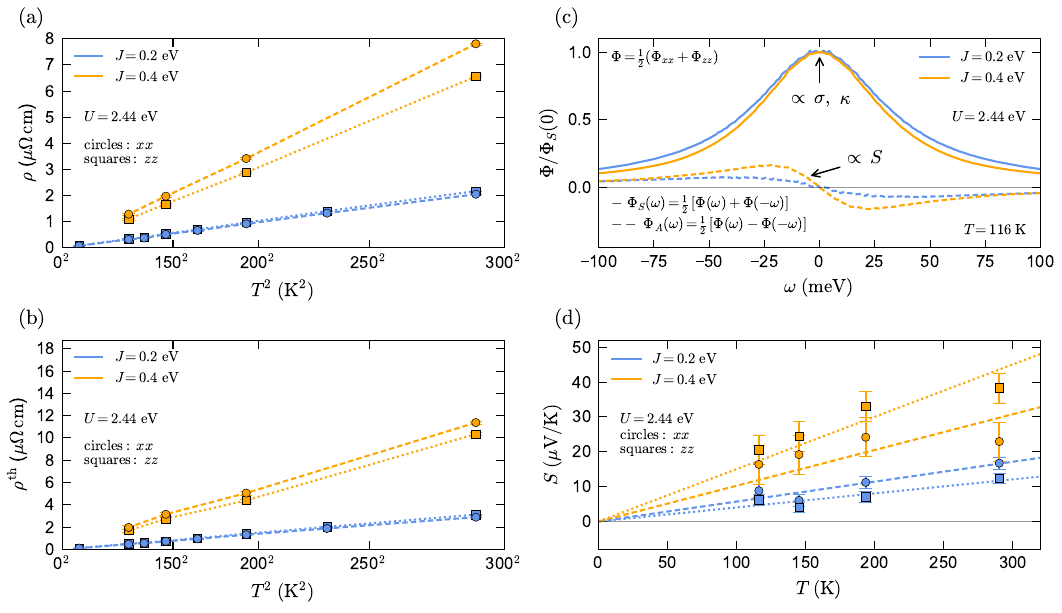}
    \caption{Calculated low-temperature transport properties of RuO$_{2}$. Temperature dependence of the (a) electrical resistivity and (b) thermal resistivity defined in Eq.~\eqref{eqn:Tth} along the $xx$ (circles) and $zz$ (squares) directions. (c) Interacting transport function (Eq.~\eqref{eqn:transp}) at $T = 116$ K decomposed into symmetric (S, solid lines) and asymmetric components (A, dashed lines) normalized to the symmetric value at zero frequency. For simplicity, we average over the in- and out-of-plane directions. (d) Temperature dependence of the Seebeck coefficient (thermopower). The error bars corrlspond to the standard deviation from a least-squares fit to the functional form $S = \alpha T$. } 
    \label{fig:transport}
\end{figure*}

\section{\label{app:extract}Extracting low-$T$ transport coefficients from DMFT data}

In this section, we describe in more detail how we extract the transport coefficients presented in Fig.~\ref{fig:exp_transport} and the raw DMFT transport data.

We compute the transport properties of RuO$_2$ within the single-site DMFT approximation over a range of electronic temperatures from 58 to 290 K. At these temperatures, electron-phonon scattering can contribute appreciably to experimentally measured transport. Our DMFT calculations instead isolate the electron-electron contribution. Because electronic correlations in RuO$_2$ are moderate, the system has an exceptionally high Fermi liquid coherence scale, allowing the electronic self-energy to exhibit Fermi liquid behavior at temperatures substantially higher than the experimentally accessible low temperature regime ($T \lesssim 20$ K). In the SM, we provide several analyses of the electronic self-energy that demonstrate the expected Fermi liquid scaling over the temperature range considered here~\cite{supp}.

\begin{figure*}
    \centering
    \includegraphics[width=2\columnwidth]{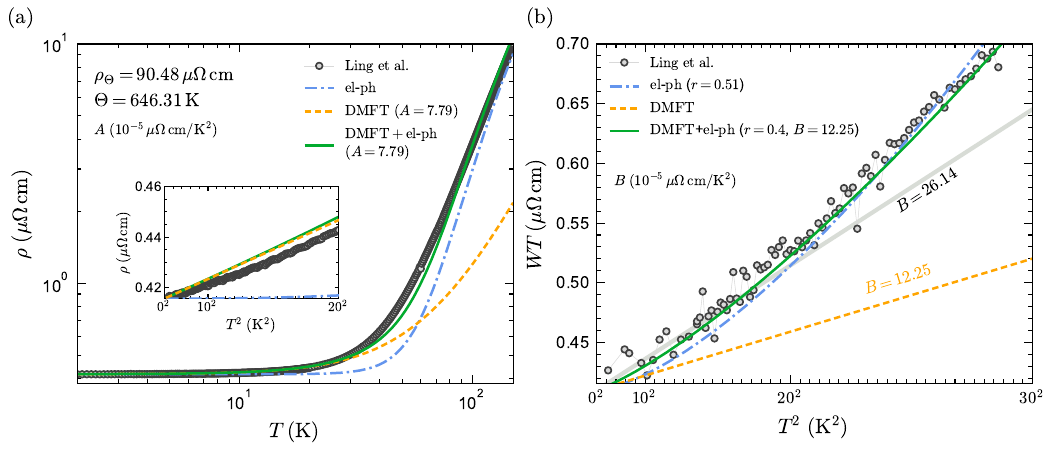}
    \caption{Derivation of the electron-phonon (el-ph) scattering model for RuO$_{2}$ obtained from the experimental data in Ref.~\onlinecite{ling2026}. (a) Log-log plot of the electrical resistivity measured along the $\vec{c}$ direction from Ref.~\onlinecite{ling2026} (sample 2). Fitting to Eq.~\eqref{eq:bloch_electrical}, we can obtain $\rho_{\Theta}$ and $\Theta$ (see text for more details). Inset: low-T blow up of $\rho$ versus $T^{2}$. (b) Thermal resistivity from Ref.~\onlinecite{ling2026} compared with three transport models: el-ph, DMFT, and DMFT+el-ph, which represent electron-phonon scattering, electron-electron-scattering, and their combination, respectively.}
    \label{fig:elph}
\end{figure*}

We therefore use the temperature dependence of the calculated transport coefficients to extract the corresponding low-temperature Fermi liquid parameters. In the Fermi liquid regime, the electrical resistivity, thermopower, and electronic thermal resistivity scale as
$\rho^{\mathrm{el}}(T)=\rho_{0}+AT^{2}$,
$S(T)=\alpha T$, and
$\rho^{\mathrm{th}}(T)=\rho_{0}^{\mathrm{th}}+BT^{2}$,
respectively. In a single-orbital or cubic system, where the self-energy is scalar valued, the electrical resistivity coefficient $A$ is directly related to the quadratic-in-frequency coefficient of the imaginary part of the electronic self-energy~\cite{kugler2026,labollita2026}. For a multi-orbital, non-cubic material such as RuO$_2$, this relationship is less direct because the self-energy and transport distribution functions are matrix valued. We therefore determine $A$, $\alpha$, and $B$ directly by fitting the calculated temperature dependence of $\rho^{\mathrm{el}}(T)$, $S(T)$, and $\rho^{\mathrm{th}}(T)$, respectively.

Panels (a) and (b) of Fig.~\ref{fig:transport} show the electrical and electronic thermal resistivities plotted as a function of $T^{2}$. Both quantities exhibit clear quadratic temperature dependence over the full temperature range considered, consistent with the expected Fermi liquid scaling. This behavior is observed for both the in-plane ($xx$) and out-of-plane ($zz$) components, although the corresponding quadratic coefficients are anisotropic and increase substantially with Hund's coupling $J$. Panel (d) shows the thermopower, which is approximately linear in temperature over the same temperature range, consistent with $S=\alpha T$.

The stronger dependence of the thermopower on $J$ can be understood from the energy dependence of the transport distribution function. Unlike the electrical and thermal conductivities, which are dominated by the particle-hole symmetric component of $\Phi(\omega)$ near the Fermi level, the thermopower is directly sensitive to its particle-hole asymmetric component. In panel (c) of Fig.~\ref{fig:transport}, we decompose the transport distribution function into symmetric and antisymmetric contributions about the Fermi level. Increasing $J$ produces a pronounced enhancement of the antisymmetric component, reflecting an increased particle-hole asymmetry of the interacting electronic structure. This naturally leads to the enhanced thermopower and larger linear coefficient $\alpha$ observed for $J=0.4$ eV.

\section{\label{app:bloch}Bloch model of electron-phonon scattering in thermal and electrical resistivity}
In this section we describe our approach to examining the role of  electron-phonon scattering in the low-temperature electronic thermal resistivity discussed in Sec.~\ref{sec:elph} and shown in Fig.~\ref{fig:exp_transport}(c). We use the Bloch model, which provides a minimal description of electron-phonon scattering in a metal. The model assumes a spherical Fermi surface with dimensionless radius $r$ and a Debye phonon characterized by the Debye temperature $\Theta$. Within this model, the electron-phonon contribution to the electronic thermal resistivity is given by~\cite{Ziman1960}
\begin{equation}
\label{eq:bloch_thermal}
\begin{split}
[WT]_{\mathrm{el-ph}}
=4\rho_{\Theta}\Bigg[
&\left(\frac{T}{\Theta}\right)^5
\left\{
\mathcal{J}_5\left(\frac{\Theta}{T}\right)
-\frac{1}{2\pi^2}
\mathcal{J}_7\left(\frac{\Theta}{T}\right)
\right\}
\\
&+
\frac{3}{\pi^2}r^2
\left(\frac{T}{\Theta}\right)^3
\mathcal{J}_5\left(\frac{\Theta}{T}\right)
\Bigg].
\end{split}
\end{equation}
where the material-dependent prefactor $\rho_{\Theta}$ has units of resistivity. The Debye functions are defined as
\begin{equation}
\mathcal{J}_n(x)
=
\int_0^x dz\,
\frac{z^n e^z}{(e^z-1)^2}.
\end{equation}
The corresponding Bloch expression for the electron-phonon contribution to the electrical resistivity is~\cite{Ziman1960}
\begin{equation}
\label{eq:bloch_electrical}
\rho_{\mathrm{el-ph}}(T)
=
4\rho_{\Theta}
\left(\frac{T}{\Theta}\right)^5
\mathcal{J}_5\left(\frac{\Theta}{T}\right).
\end{equation}

An important feature of the Bloch model is the different low-temperature scaling of the electron-phonon contributions to charge and heat transport. For $T\ll\Theta$, the electrical resistivity varies as $T^5$, while the leading electron-phonon contribution to the electronic thermal resistivity varies as $T^3$. Electron-phonon scattering can therefore be negligible in the electrical resistivity while still providing an appreciable contribution to the relaxation of the electronic heat current. This difference is central to the interpretation of the thermal-transport discrepancy considered here.

We first determine $\Theta$ and $\rho_{\Theta}$ from the measured electrical resistivity using Eq.~\eqref{eq:bloch_electrical}. Panel (a) of Fig.~\ref{fig:elph} shows the electrical resistivity of sample 2 from Ref.~\onlinecite{ling2026}, together with the corresponding low-temperature fits. We obtain $\rho_{\Theta}=90.5~\mu\Omega\,\mathrm{cm}$ and $\Theta=646~\mathrm{K}$. The low-temperature inset shows that the electron-phonon contribution to the electrical resistivity is negligible in the Fermi liquid regime, where the measured resistivity is instead dominated by electron-electron scattering.

With $\rho_{\Theta}$ and $\Theta$ fixed by the electrical transport data, the only remaining parameter entering the Bloch model for the electronic thermal resistivity is the dimensionless Fermi-surface radius $r$. Panel (b) of Fig.~\ref{fig:elph} shows the measured low-temperature electronic thermal resistivity of sample 2 from Ref.~\onlinecite{ling2026}. As discussed in the main text, the experimental thermal resistivity cannot be quantitatively described by the electron-electron contribution obtained from DMFT alone. If the thermal resistivity is modeled using only the electron-phonon contribution, the overall magnitude of the experimental data can be reproduced with $r=0.51$. This demonstrates that electron-phonon scattering can naturally generate a thermal resistivity of the required scale.

A better description is obtained when the independently calculated DMFT electron-electron contribution is combined with the Bloch electron-phonon contribution. In this case, the best comparison with experiment is obtained for $r=0.40$. This analysis therefore provides a plausible explanation for why the experimentally inferred thermal resistivity differs substantially from the DMFT result, while the electrical resistivity remains well described. 

We emphasize that the Bloch analysis is phenomenological and does not constitute a microscopic calculation of electron-phonon heat-current relaxation in RuO$_2$. Rather, it demonstrates that a low-temperature electron-phonon contribution can significantly complicate the extraction of the intrinsic electron-electron thermal resistivity. A quantitative resolution of the heat-transport puzzle identified in this work will therefore require a microscopic treatment of both electron-electron and electron-phonon scattering processes.

\bibliography{ref.bib}

\end{document}

\title{Supplemental Material to\\ ``Unraveling Fermid liquid charge and heat transport in RuO$_{2}$''}
\author{Harrison LaBollita\orcidlink{0000-0002-6699-8577}}
\email{hlabollita@flatironinstitute.org}
\affiliation{Center for Computational Quantum Physics, Flatiron Institute, New York, New York 10010, USA}             
\author{Fabian B.~Kugler\orcidlink{0000-0002-3108-6607}}
\affiliation{Institute for Theoretical Physics, University of Cologne, 50937 Cologne, Germany}
\author{Antoine Georges\orcidlink{0000-0001-9479-9682}}
\affiliation{Coll{\`e}ge de France, 11 Place Marcelin Berthelot, 75005 Paris, France}
\affiliation{Center for Computational Quantum Physics, Flatiron Institute, New York, New York 10010, USA}             
\affiliation{CPHT, CNRS, {\'E}cole Polytechnique, IP Paris, F-91128 Palaiseau, France}
\affiliation{DQMP, Universit{\'e} de Gen{\`e}ve, 24 Quai Ernest Ansermet, CH-1211 Gen{\`e}ve, Suisse}
\author{Cyrus E. Dreyer\orcidlink{0000-0003-0035-7520}}
\affiliation{Department of Physics and Astronomy,
             Stony Brook University,
             Stony Brook, New York, 11794-3800, USA}
\affiliation{Center for Computational Quantum Physics,
             Flatiron Institute,
             162 5th Avenue, New York, New York 10010, USA.}
\affiliation{Center for Computational Quantum Physics, Flatiron Institute, New York, New York 10010, USA}

\maketitle

This Supplemental Material provides additional details supporting the calculations and analysis presented in the main text. We first summarize the transport formalism used to compute the optical and dc electrical, thermoelectric, and thermal transport coefficients within DMFT. We then present additional DMFT results for RuO$_2$, including the dependence of the electronic self-energies and local observables on Hund's coupling, as well as the Fermi-liquid scaling of both the self-energies and the interacting transport function.  Finally, we provide additional numerical checks of the transport calculations, including convergence with respect to Brillouin-zone sampling and the influence of the static real part of the DMFT self-energy on the calculated resistivities.

\tableofcontents

\section{\label{sec:formalism}Transport formalism within the dynamical mean-field approximation}
In this section, we review the transport formalism within the dynamical mean-field (DMFT) approximation. Within DMFT, neglecting vertex corrections, the optical conductivity simplifies to 
\begin{equation}
\label{eqn:optics}
\sigma_{ab}(\Omega) = 2\pi e^{2}\hbar\int d\omega \, \frac{f(\omega) - f(\omega + \Omega)}{\Omega}\Phi_{ab}(\omega, \Omega),
\end{equation}
where $f(\omega)=(1+e^{\beta\omega})^{-1}$ is the Fermi function and $\beta = 1/k_{\mathrm{B}}T$. The transport distribution function $\Phi_{ab}(\omega, \Omega)$ is defined as 
\begin{equation}
\label{eqn:transp}
\Phi_{ab}(\omega, \Omega)= \frac{1}{V} \sum_{\vec{k}}\mathrm{Tr}[v_{\vec{k}}^{a} A_{\vec{k}}(\omega)v_{\vec{k}}^{b}A_{\vec{k}}(\omega+\Omega)],
\end{equation}
$v_{\vec{k}}^{a} = \frac{1}{\hbar}\partial_{k_{\alpha}}H_{\vec{k}} +\frac{1}{\hbar} i[H_{\vec{k}},R_{\vec{k}}^{a}]$
are the velocity matrix elements along Cartesian direction $a$ ($R_{\vec{k}}^{a}= \langle\phi_{m\vec{k}}|\hat{r}_{a}|\phi_{\vec{k}m'}\rangle$) and
$A_{\vec{k}}(\omega) = \frac{1}{2\pi i}[G_{\vec{k}}(\omega)-G^{\dagger}_{\vec{k}}(\omega)]$
is the interacting spectral function with $G_{\vec{k}}(\omega) = [\omega + \mu -H_{\vec{k}} - \Sigma(\omega)]^{-1}$. Note $v_{\vec{k}}^{a}$ and $A_{\vec{k}}(\omega)$ are matrices with their dimension corresponding to the number of orbitals. 

Taking the $\Omega=0$ limit, we obtain the dc electrical conductivity. The thermopower and electronic thermal conductivity (respectively) are obtained from energy-weighted moments of the dc transport distribution $\Phi_{ab}(\omega)\equiv\Phi_{ab}(\omega, 0)$. For a tetragonal system:

\begin{align}
\label{eqn:kinetic}
    \sigma_{aa} &= \beta e^{2} [A_{0}]_{aa},\\
    S_{aa} &= -\frac{k_{\rm B}}{|e|}\frac{[A_{1}]_{aa}}{[A_{0}]_{aa}},\\ [\kappa_{e}]_{aa} &= k_{\mathrm{B}}[[A_{2}]_{aa} - ([A_{1}]_{aa})^{2}/[A_{0}]_{aa}],
\end{align}
from the kinetic coefficients which are defined as 
\begin{equation}
    [A_{n}]_{aa} = 2\pi\hbar \int d\omega\, (\beta\omega)^{n}f(\omega)(1-f(\omega))\Phi_{aa}(\omega),
\end{equation}
where $\Phi_{aa}(\omega) = \Phi_{aa}(\omega, \Omega =0)$. The dc electrical resistivity is the inverse of the dc electrical conductivity 
\begin{equation}
    \label{eqn:Tel}
\rho^{\text{el}}_{aa} = [\sigma^{-1}]_{aa}.
\end{equation}
To facilitate a direct comparison between the electrical resistivity and the electronic contribution to the thermal resistivity, the latter is often expressed as,
\begin{equation}
\label{eqn:Tth}
\rho^{\mathrm{th}}_{aa} = L_{0}T[\kappa_{e}^{-1}]_{aa},
\end{equation}
with $L_{0} = (k_{\rm B}^{2}/e^{2}) \pi^{2}/3$, the Sommerfeld value of the Lorenz ratio. Experimentally, this definition establishes consistent units between the resistivities and ensures the residual $T=0$ thermal resistivity is the same as the electrical resistivity [$\rho^{\mathrm{th}}(0)=\rho^{\mathrm{el}}(0)$].

\section{\label{sec:adddmft}Additional DMFT data}
In this section, we provide additional data from our dynamical mean-field calculations for RuO$_{2}$ including an an analysis of the Hund's coupling, Fermi-liquid collapse of the electronic self-energies, and the corresponding Fermi-liquid collapse of the DMFT transport function.

\subsection{\label{sec:hunds}Hund's coupling analysis}

\begin{figure*}
   \includegraphics[width=\columnwidth]{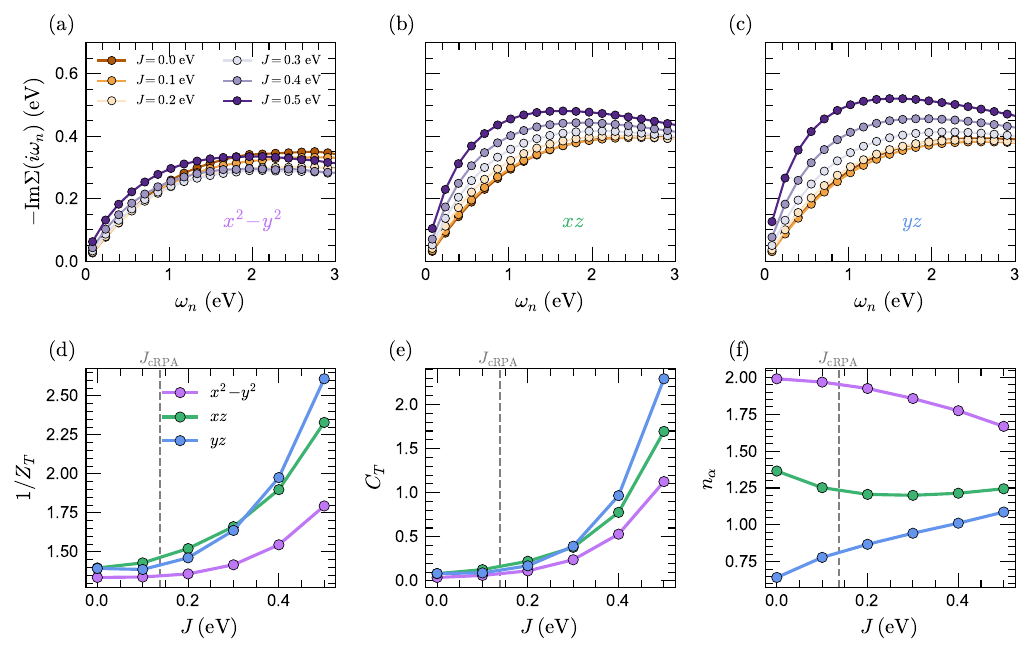} 
   \caption{Effects of Hund's coupling ($J$) on local DMFT self-energies and observables with $U=U_{\mathrm{cRPA}}=2.44$ eV at $T = 1/40$ eV ($\simeq290$ K). Imaginary part of the DMFT Matsubara self-energies as a function of $J$ for the (a) $x^{2}-y^{2}$, (b) $xz$ and (c) $yz$ Wannier orbitals. Orbitally-resolved (d) $1/Z_{T}$, (e) $C_{T}$, and (f) occupations as a function of $J$. Dashed line (grey) in (d-f) denotes $J_{\mathrm{cRPA}}$.
   %\fk{Increase font size of legend a bit.}\hl{done}
   }
   \label{fig:hunds}
\end{figure*}

Figure~\ref{fig:hunds} summarizes the effects of Hund's coupling on the DMFT self-energies and local observables for RuO$_{2}$ at $T = 1/40$ eV ($\simeq290$ K). We fix the intraorbital Coulomb interaction to the cRPA value of $2.44$ eV. The interorbital Coulomb interaction is fixed to $U^{\prime} = U - 2J$. The orbitally-resolved DMFT self-energies are shown in Fig.~\ref{fig:hunds}(a-c) for the Ru-$\{x^{2}-y^{2},xz,yz\}$, respectively. The effects of Hund's coupling increases the effects the strength of correlations for the $xz$ and $yz$ orbitals (increasing slope of $-\mathrm{Im}\Sigma(i\omega_{n})$ at low frequencies). In terms of local observables, we track the effects of Hund's coupling on $1/\mathcal{Z}_{T}$, $\mathcal{C}_{T}$, and the orbital occupations ($n_{\alpha}$). We have used the following definitions for $1/Z_{T}$ and $C_{T}$ as discussed in Ref.~\onlinecite{labollita2026}, 
\begin{equation}
\label{eqn:s_iwn}
1-\frac{1}{\mathcal{Z}_{T}} = \frac{\mathrm{Im}\Sigma(i\pi T)}{\pi T},\quad \mathcal{C}_{T} = \frac{\mathrm{Im}\Sigma(3i\pi T)- 3\mathrm{Im}\Sigma(i\pi T)}{8\pi^{2}T^{2}}.
\end{equation}
These local observables are summarized in Fig.~\ref{fig:hunds}(d-f). Increasing Hund's coupling can drastically increase the mass renormalizations compared to the band mass in all orbital sectors as shown in Fig.~\ref{fig:hunds}(d), as well as the quadratic coefficient of the self-energy, as defined in Eq.~\eqref{eqn:s_iwn} (see Fig.~\ref{fig:hunds}(e)). As discussed in the main text, using the cRPA values to parameterize the screened Coulomb interaction underestimate the specific-heat coefficient determined from experiments. Increasing the Hund's coupling improves the overall agreement with experiment. Finally, from the orbital-resolved occupations ($n_{\alpha}$) computed from the local Green's function on the Mastubara axis shows how the Hund's coupling drives the $xz$ and $yz$ orbitals closer to half-filling (see Fig.~\ref{fig:hunds}(f)).

\subsection{\label{sec:collapse}Fermi-liquid scaling collapse of DMFT self-energies $\Sigma(\omega, T)$}
In the FL regime, the local self-energy has the low-frequency
form
\begin{equation}
-\mathrm{Im}\,\Sigma_m(\omega,T)
=
C_m\left[\omega^2+(\pi T)^2\right],
\label{eq:fl_self_energy}
\end{equation}
where $m$ denotes the orbital index. The coefficient $C_m$ measures the strength of quasiparticle scattering in orbital $m$. Equivalently,
\begin{equation}
-\frac{\mathrm{Im}\,\Sigma_m(\omega,T)}{(\pi T)^2}
=
C_m\left[1+\left(\frac{\omega}{\pi T}\right)^2\right],
\end{equation}
so that different temperatures collapse onto a common scaling function
when plotted as a function of $|\omega|/T$.  

\begin{figure*}
    \includegraphics[width=\columnwidth]{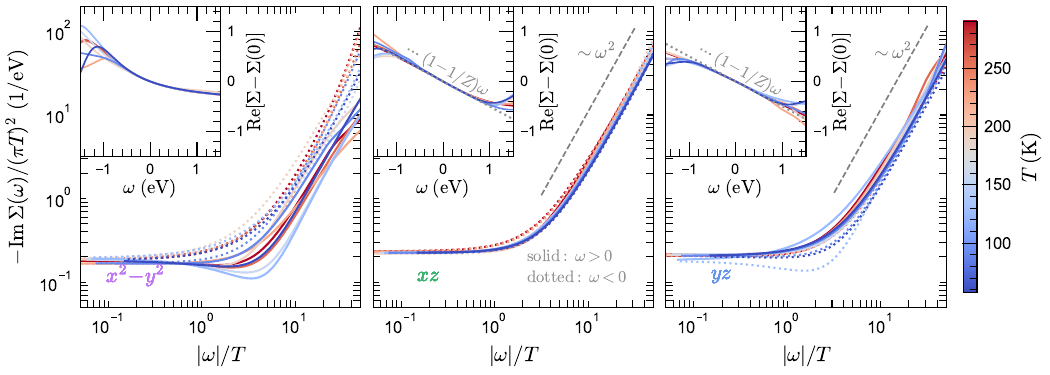}
    \caption{Log-log plot of $-\mathrm{Im}\Sigma(\omega, T)/(\pi T)^{2}$ versus $|\omega|/T$ for the three Wannier orbitals of Ru: $x^{2}-y^{2}$, $xz$, $yz$ (left to right). Solid (dotted) lines correspond to $\omega > 0$ ($\omega < 0$). Grey dashed lines indicate $\omega^{2}$ scaling. Inset: $\mathrm{Re}[\Sigma(\omega, T)-\Sigma(0, T)]$ compared with the low-frequency slope of $1-1/Z = \partial_{\omega}\mathrm{Re}\Sigma(\omega, T)|_{\omega=0}$. The $xz$ and $yz$ orbitals collapse onto the expected FL form $C(1+\omega^{2}/(\pi T)^{2})$, while the $x^{2}-y^{2}$ shows deviations. The interaction parameters used were $U = 2.44$ eV and $J =0.2$ eV.}  
    \label{fig:flw}
\end{figure*}

\begin{figure*}
    \includegraphics[width=\columnwidth]{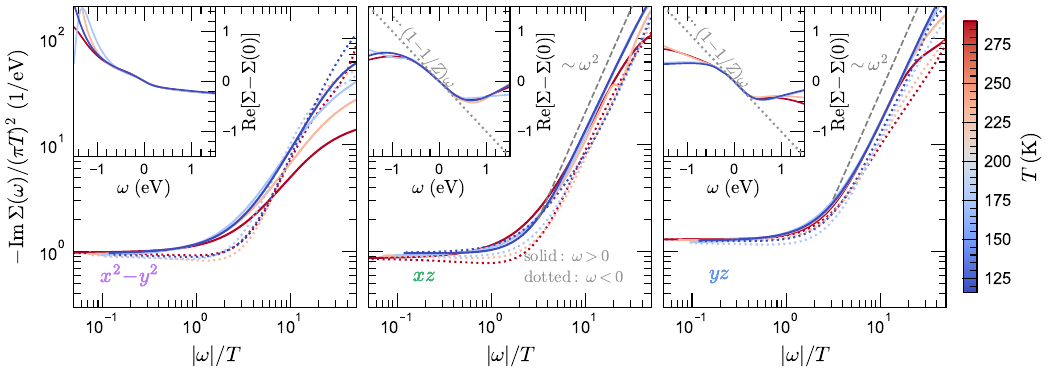}
    \caption{Same as Fig.~\ref{fig:flw} with $U=2.44$ eV and $J=0.4$ eV.}  
    \label{fig:flw2}
\end{figure*}

Figure~\ref{fig:flw} shows the FL scaling collapse (orbitally-resolved) $-\mathrm{Im}\Sigma_{m}(\omega)/(\pi T)^{2}$ for RuO$_{2}$ using $U = 2.44$ eV and $J = 0.2$ eV to parameterize the Hubbard-Kanamori Hamiltonian. From the data, we can see that the quasiparticle renormalization $Z$wdecrease for each orbital, as well as the quadratic coefficient of the self-energy increases in line with our expectations from a similar analysis on the Matsubara axis shown in Fig.~\ref{fig:hunds}.  The same data is provlded for the self-energies obtained with a larger value of Hund's coupling $J =0.4$ eV as shown in Fig.~\ref{fig:flw2}. For both datasets, we are able to obtain an (approximate) FL scaling collapse of our DMFT self-energies indicating that our calculations are within the FL regime of RuO$_{2}$. This allows us to make statements about the low-temperature FL properties of RuO$_{2}$ while remaining at relatively high electronic temperatures for the numerical calculations.

\subsection{\label{sec:dmfttransport}Transport function $\Phi(\omega, T)$ and scaling collapse}
\begin{figure*}
    \centering
    \includegraphics[width=\columnwidth]{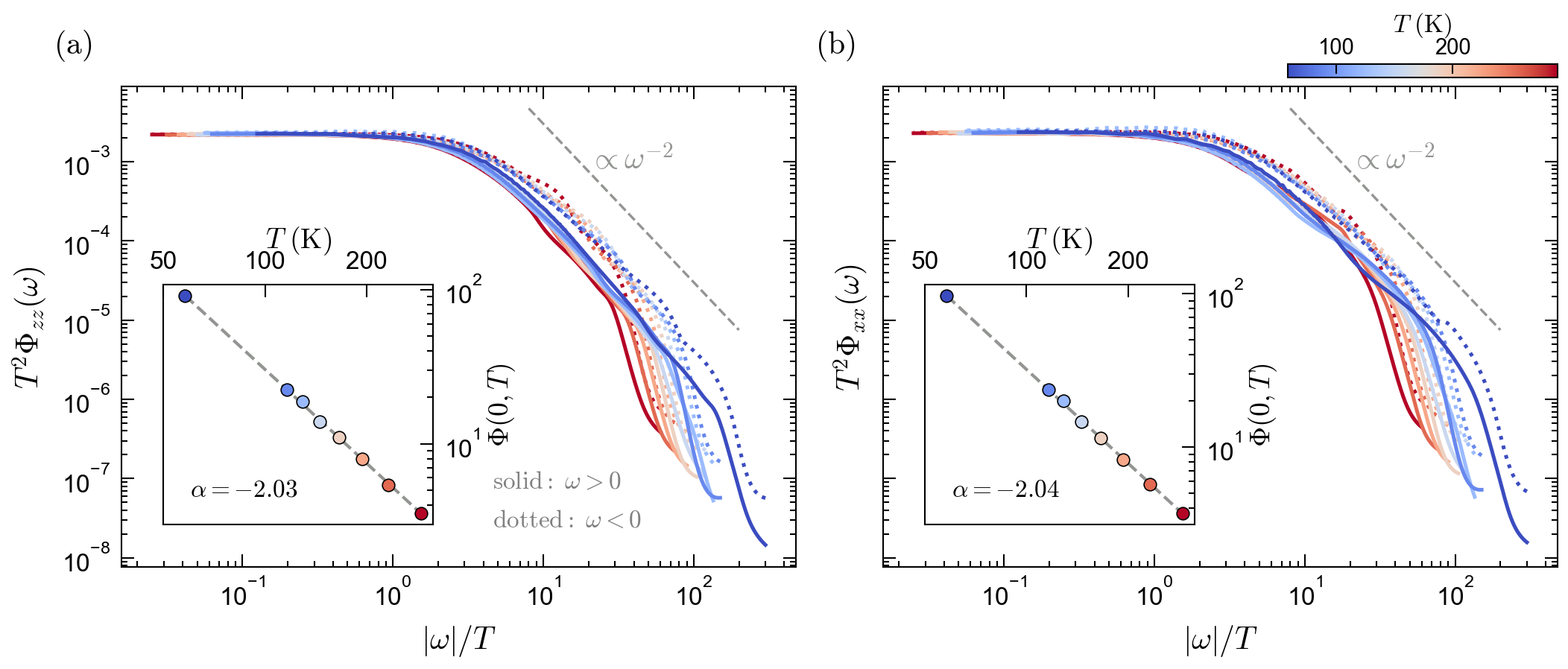}
    \caption{Log-log plot the DMFT transport function for RuO$_{2}$ for $U = 2.44$ eV and $J=0.2$ eV showing the analogous FL collapse as the self-energy. Insets: $\Phi_{aa}(\omega=0,T)$ as a function of $T$ on a log-line scale. (a) $zz$ direction. (b) $xx$ direction.}
    \label{fig:dmfttransport}
\end{figure*}

The interacting transport function is defined in Eq.~\ref{eqn:transp}. For a scalar-valued self-energy, appropriate for single-orbital or cubic materials, implies $\Phi_{aa}(\omega,T) \propto |\mathrm{Im}\Sigma(\omega,T)|^{-1} \Phi^{U=0}_{aa}(\varepsilon_{\mathrm{F}})$~\cite{kugler2026,labollita2026}, where $\Phi^{U=0}_{aa}(\varepsilon_{\rm F})$ is the non-interacting transport function. Generalizing to the multi-orbital, matrix-valued case, which is relevant for RuO$_{2}$, we have,
\begin{equation}
\Phi_{aa}(\omega,T)
=
T^{-2}\varphi_{a}(|\omega|/T),
\label{eq:phi_scaling}
\end{equation}
where $\varphi_{a}(|\omega|/T)$ is a generic function of $|\omega|/T$.  Figure~\ref{fig:dmfttransport} shows that the DMFT transport distribution function (Eq.~\eqref{eqn:transp}) obeys this scaling form, with $\Phi_{aa}(0)\sim T^{-2}$. 

\section{Additional details of the transport calculations}
\subsection{Convergence of Brillouin zone sampling}
We examine the convergence of the DC resistivity tensor with respect to the Brillouin zone sampling. The conductivity is evaluated on uniform $k$-point girds containing $N_{k}^{3}$ points, and the corresponding resistivity components are obtained from the inverse of the conductivity tensor.

\begin{figure}
    \centering
    \includegraphics[width=0.5\columnwidth]{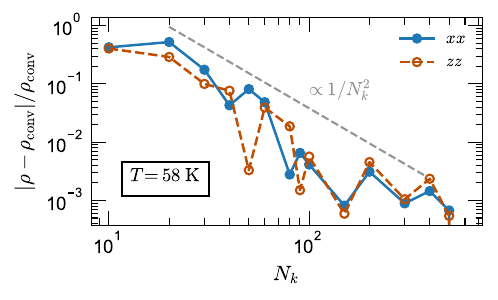}
    \caption{Convergence of the DC resistivity with respect to the Brillouin zone sampling $N_{k}$ for RuO$_{2}$ at $T=1/200$ eV ($\simeq 58$ K). Relative deviation from the value at the finest grid, $|\rho-\rho_{\mathrm{conv}}|$, on a log-log scale.}
    \label{fig:sigdc_conv}
\end{figure}

Figure~\ref{fig:sigdc_conv} shows the convergence of $\rho_{\alpha\alpha}$ as a function of the linear grid size $N_{k}$ for RuO$_{2}$ with $\Sigma(\omega)$ obtained at $T=1/200$ eV ($\simeq58$ K). We quantify the numerical error by the relative deviation from the result obtained on the finest grid, $|\rho(N_{k})-\rho(N_{k}^{\mathrm{max}})|/\rho(N_{k}^{\mathrm{max}})$, which provides a dimensionless measure of the convergence. Both components exhibit a systematic decrease of the error with increasing $k$-point density. For grid sizes used in the main text, the relative error is below $\mathcal{O}(10^{-3})$, demonstrating that the transport coefficients are well converged with respect to momentum space resolution.

\subsection{Effects of $\mathrm{Re}\Sigma(\omega=0)$ on resistivity}
\begin{figure*}
    \centering
    \includegraphics[width=0.9\columnwidth]{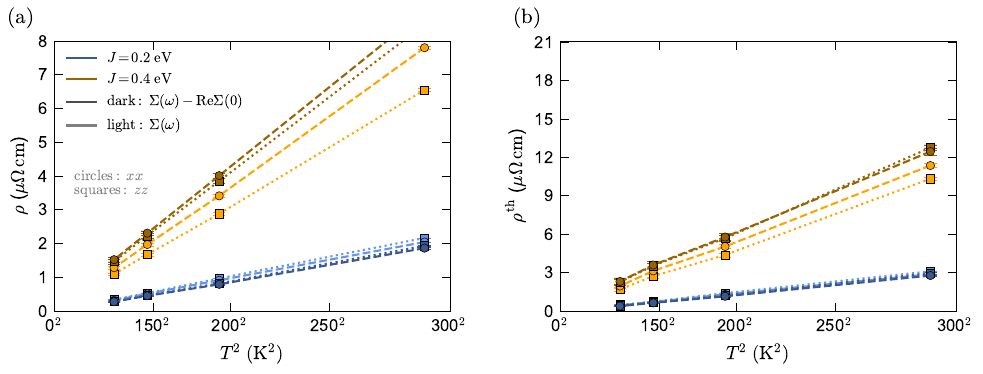}
    \caption{Transport properties of RuO$_{2}$ using $\tilde{\Sigma}(\omega) = \Sigma(\omega) - \mathrm{Re}\Sigma(0)$ from DMFT with different choices of Hund's coupling. (a) Electrical and (b) thermal resistivity along the two inequivalent Cartesian directions.}
    \label{fig:rho_resigma0}
\end{figure*}

We investigate the crystal field renormalization effects on our transport calculations by removing $\mathrm{Re}\Sigma(\omega=0)$ from the DMFT self-energy. For Sr$_{2}$RuO$_{4}$, this was shown to lead to a better agreement with angle-resolved photoemission data~\cite{Tamai2019}. 
Because the transport integrals are sensitive to details of the Fermi surface, it is important to understand crystal field renormalization effects. To that end, we perform a similar analysis for RuO$_{2}$. We replace the DMFT self-energy, $\Sigma(\omega)$, with $\tilde{\Sigma}(\omega) = \Sigma(\omega) -\mathrm{Re}\Sigma(0)$ and set $\mu = \varepsilon_{\rm F}$. This subtraction removes the DMFT crystal field renormalization encoded in the zero frequency self-energy. We compute the transport properties using the same formalism described in Sec.~\ref{sec:formalism} with $\tilde{\Sigma}(\omega)$. Our results are summarized in Fig.~\ref{fig:rho_resigma0}. We find thatwremoving $\mathrm{Re}\Sigma(0)$ from the self-energy results in an increase in both the electrical and thermal resistivity. This increase scales with the size of the Hund's coupling meaning a larger increase in resistivity for a larger Hund's coupling. However, this approximated DMFT self-energy does not lead to a noticeably better agreement with the available experimental data for either the electrical or thermal transport~\cite{ling2026}. Detailed photoemission studies will help elucidate details of the crystal field renormalizations.

\bibliography{ref}